\documentclass[italian,english]{article}
\usepackage[latin9]{inputenc}
\usepackage{color}
\usepackage{amssymb}

\newcommand{\lyxaddress}[1]{
\par {\raggedright #1
\vspace{1.4em}
\noindent\par}
}

\usepackage{babel}

\begin{document}

\title{\textbf{Primordial inflation from gravity's rainbow}}

\author{\textbf{Christian Corda }}

\maketitle

\lyxaddress{\begin{center}
Associazione Scientifica Galileo Galilei, Via Bruno Buozzi 47 - 59100
PRATO, Italy 
\par\end{center}}

\begin{center}
\textit{E-mail address:} \textcolor{blue}{cordac.galilei@gmail.com}
\par\end{center}
\begin{abstract}
In a recent paper, which has been published in Nature, the LIGO Scientific
Collaboration (LSC) obtained an upper limit on the stochastic gravitational-wave
background of cosmological origin by using the data from a two-year
science run of the Laser Interferometer Gravitational-wave Observatory
(LIGO). Such an upper limit rules out some models of early Universe
evolution, like the ones with relatively large equation-of-state parameter
and the cosmic (super) string models with relatively small string
tension arising from some String Theory's models. It results also
an upper limit for the relic stochastic background of gravitational
waves (RSBGWs) which is proposed by the Pre-Big-Bang Theory.

On the other hand, the upper bound on the RSBGWs which is proposed
by the Standard Inflationary Model is well known and often updated
by using the Wilkinson Microwave Anisotropy Probe (WMAP).

The potential detection of such a RSBGWs is the only way to learn
about the evolution of the very early universe, up to the bounds of
the Planck epoch and the initial singularity. This is a kind of information
that is inaccessible to standard astrophysical observations.

By using a conformal treatment, a formula that directly connects the
average amplitude of the RSBGWs with the Inflaton field has been recently
obtained in our paper Gen. Rel. Grav. 42, 5, 1323-1333 (2010). In
this proceeding, by joining this formula with the equation for the
characteristic amplitude $h_{C}$ for the RSBGWs, the upper bounds
on the RSBGWs from the WMAP and LSC data will be translated in lower
bounds on the Inflaton field.

The results show that the value of the Inflaton field that arises
from the WMAP bound on the RSBGWs is totally in agreement with the
famous \textit{slow roll} condition on Inflation, while the value
of the Inflaton field that arises from the LSC bound on the RSBGWs
could be not in agreement with such a condition. 
\end{abstract}

\section{Introduction}

The scientific community aims in a first direct detection of GWs in
next years (for the current status of GWs interferometers see \cite{key-1})
confirming the indirect, Nobel Prize Winner, proof of Hulse and Taylor
\cite{key-2}.

Detectors for GWs will be important for a better knowledge of the
Universe and either to confirm or to rule out, in an ultimate way,
the physical consistency of General Relativity, eventually becoming
an observable endorsement of Extended Theories of Gravity \cite{key-3}.

It is well known that an important potential source of gravitational
radiation is the relic stochastic background of GWs \cite{key-4}.
The potential existence of such a relic stochastic background arises
from general assumptions that mix principles of classical gravity
and principles of quantum field theory \foreignlanguage{italian}{\cite{key-5,key-6,key-7}}.
As the zero-point quantum oscillations, which produce relic GWs, are
generated by strong variations of the gravitational field in the early
universe, the potential detection of relic GWs is the only way to
learn about the evolution of the very early universe, up to the bounds
of the Planck epoch and the initial singularity \cite{key-4,key-7}.
The importance of this \textit{gravity's rainbow} in cosmological
scenarios has been discussed in an elegant way in \cite{key-8}.

The inflationary scenario for the early univers\foreignlanguage{italian}{e
\cite{key-9,key-10}}, which is tuned in a good way with the WMAP
data on the Cosmic Background Radiation (CBR) (in particular exponential
inflation and spectral index $\approx1$ \foreignlanguage{italian}{\cite{key-11})}
amplified the zero-point quantum oscillations \cite{key-6,key-7}.

A recent paper, which has been written by the LSC \cite{key-4}, has
shown an upper limit on the RSBGWs by using the data from a two-year
science run of LIGO. Such an upper limit rules out some models of
early Universe evolution, like the ones with relatively large equation-of-state
parameter and the cosmic (super) string models with relatively small
string tension arising from some string theory models. It results
also an upper limit for the RSBGWs which is proposed by the Pre-Big-Bang
Theory (see \cite{key-4} for details).

Another well known upper bound on the RSBGWs arises from the Standard
Inflationary Model. Such an upper bound is often updated by using
the WMAP data \cite{key-4,key-12}. 

It is well known that the potential detection of such a RSBGWs is
the only way to learn about the evolution of the very early universe,
up to the bounds of the Planck epoch and the initial singularity.
In fact, this kind of information is inaccessible to standard astrophysical
observations \cite{key-4,key-7,key-8}.

In this proceeding, a formula that directly connects the average amplitude
of the RSBGWs with the Inflaton field, that has been obtained in \cite{key-13},
will be used, together with the equation for the characteristic amplitude
$h_{C}$ for the RSBGWs \cite{key-15}, in order to translate the
upper bounds on the RSBGWs from the WMAP and LSC data in lower bounds
on the Inflaton field.

Our results show that the value of the Inflaton field that arises
from the WMAP bound on the RSBGWs is totally in agreement with the
famous \textit{slow roll} condition on Inflation \foreignlanguage{italian}{\cite{key-9,key-10}},
while the value of the Inflaton field that arises from the LSC bound
on the RSBGWs could not be in agreement with this condition.

\section{The spectrum of the relic gravitational waves}

Considering a stochastic background of GWs, it can be characterized
by a dimensionless spectrum \cite{key-4,key-7,key-8}. The more recent
values for the spectrum that arises from the WMAP data can be found
in refs. \cite{key-4,key-12}. In such papers it is (for a sake of
simplicity, in this paper natural units are used, i.e. $8\pi G=1$,
$c=1$ and $\hbar=1$) \begin{equation}
\Omega_{gw}(f)\equiv\frac{1}{\rho_{c}}\frac{d\rho_{gw}}{d\ln f}\leq10^{-13}\label{eq: spettro}\end{equation}

where \begin{equation}
\rho_{c}\equiv3H_{0}^{2}\label{eq: densita critica}\end{equation}

is the (actual) critical density energy, $\rho_{c}$ of the Universe,
$H_{0}$ the actual value of the Hubble expansion rate and $d\rho_{gw}$
the energy density of relic GWs in the frequency range $f$ to $f+df$.
This is the upper bound on the RSBGWs that observations put on the
Standard Inflationary Model.

An higher bound results from the LIGO Scientific Community data in
ref. \cite{key-4}:

\begin{equation}
\Omega_{gw}\leq6.9*10^{-6}.\label{eq: spettro LIGO}\end{equation}

This bound is at $95\%$ confidence in the frequency band $41.5-169.25Hz$,
(see \cite{key-4} for details).

In this case, the value is an upper limit for the RSBGWs which arises
from the Pre-Big-Bang Theory \cite{key-4,key-16}. It also rules out
some models of early Universe evolution, like the ones with relatively
large equation of state parameter and the cosmic (super) string models
with relatively small string tension arising from some string theory
models (see \cite{key-4}and references within).

\section{Bounds from observations}

Let us consider the computation in \cite{key-13}. In such a paper
a conformal treatment has been used to obtain:

\begin{equation}
\varphi=\frac{H^{2}}{2A_{h}^{2}}\label{eq: fi}\end{equation}

(see Equation 42 in \cite{key-13}), where $\varphi$ is the Inflaton
field which generates Inflation, $H$ the value of the Hubble expansion
rate at the first horizon crossing and the averaged amplitude $A_{h}$
of the perturbations of the RSBGWs is defined like \begin{equation}
A_{h}\equiv(k/2\pi)^{3/2}h\label{eq: Ah}\end{equation}

(see \cite{key-13}). 

We emphasize the importance of the formula (\ref{eq: fi}). If the
GWs interferometers will detect the RSBGWs in next years, such a formula
will permit to directly compute the amount of Inflation in the early
Universe. A similar computation was also performed in \cite{key-14}
in the framework of $f(R)$ Theories of Gravity.

The equation for the characteristic amplitude $h_{C}$ is (see Equation
65 in \cite{key-15})

\begin{equation}
h_{C}(f)\simeq1.26*10^{-18}\left(\frac{1{\rm Hz}}{f}\right)\sqrt{h_{100}^{2}\Omega_{gw}(f)},\label{eq: legame ampiezza-spettro}\end{equation}

where $h_{100}\simeq0.71$ is the best-fit value on the Hubble constant
\cite{key-11}. This equation gives a value of the amplitude of the
relic GWs stochastic background in function of the spectrum in the
frequency range of ground based detectors \cite{key-15}. Such an
amplitude is also the averaged strain applied on the detector's arms
by the RSBGWs \cite{key-15}. Such a range is given by the interval
$10Hz\leq f\leq10KHz$ \cite{key-1}.

Defining the average value of $h_{C}(f)$ like

\begin{equation}
A_{hc}\equiv\frac{\int1.26*10^{-18}\sqrt{h_{100}^{2}\Omega_{gw}(f)}f^{-1}df}{\int df}\label{eq:average}\end{equation}

one can assume that it is $A_{hc}\simeq A_{h}$ \cite{key-13}. 

In this way, from the fundamental eq. (\ref{eq: fi}), it is also 

\begin{equation}
\varphi\simeq\frac{H^{2}}{2A_{hc}^{2}}.\label{eq: fi 2}\end{equation}

Now, by using eq. (\ref{eq: fi 2}), we can use the bounds (\ref{eq: spettro})
and (\ref{eq: spettro LIGO}) on the RSBGWs in order to obtain bounds
on the Inflaton field $\varphi.$ First of all, we emphasize that
a redshift correction is needed because $H$ in eq. (\ref{eq: fi 2})
is computed at the time of the first horizon crossing, while the value
of $A_{hc}$ from the WMAP and LSC data is computed at the present
time of the cosmological Era. The redshift correction on the spectrum
is well known \cite{key-7}:

\begin{equation}
\Omega_{gw}(f)=\Omega_{gw}^{0}(f)(1+z_{eq})^{-1},\label{eq: correzione}\end{equation}

where $\Omega_{gw}^{0}(f)$ is the value of the spectrum at the first
horizon crossing and $z_{eq}\simeq3200$ \cite{key-11} is the redshift
of the Universe when the matter and radiation energy density were
equal, see \cite{key-7} for details. 

Then, eq. (\ref{eq: fi 2}) becomes

\begin{equation}
\varphi\simeq\frac{H^{2}}{2A_{hc}^{2}(1+z_{eq})}.\label{eq: fi 3}\end{equation}

By considering the WMAP bound (\ref{eq: spettro}), the integrals
in eq. (\ref{eq:average}) has to be computed in the frequency range
of ground based detectors which is the interval $10Hz\leq f\leq10KHz$.
One gets $A_{hc}^{2}\simeq10^{-51}$.

By restoring ordinary units and recalling that $H\simeq10^{22}Hz$
at the first horizon crossing \cite{key-7}, at the end, from eq.
(\ref{eq: fi 3}), it is \begin{equation}
\varphi\geq10^{2}grams.\label{eq: inflaton value}\end{equation}

This result represents a lower bound for the value of the Inflaton
field that arises from the WMAP data on the RSBGWs in the case of
Standard Inflation \cite{key-4,key-12}. 

Now, let us consider the LSC bound (\ref{eq: spettro LIGO}). Such
a bound is at $95\%$ confidence in the frequency band $41.5-169.25Hz$
\cite{key-4}, thus, in principle, we could not extend the integrals
in eq. (\ref{eq:average}) to the total interval $10Hz\leq f\leq10KHz$.
However, it is well known that for frequencies that are smaller than
some hertz the spectrum which arises from the Pre-Big-Bang Theory
rapidly falls, while at higher frequencies the spectrum is almost
flat with a small decreasing \cite{key-4,key-16}. Thus, the integration
of eq. (\ref{eq:average}) in the interval $10Hz\leq f\leq10KHz$
gives a solid upper bound for $A_{hc}$ in these models. One gets
$A_{hc}^{2}\simeq10^{-44}$. In this case, by restoring ordinary units
and putting the value $H\simeq10^{22}Hz$ in eq. (\ref{eq: fi 3})
it is \begin{equation}
\varphi\geq10^{-5}grams.\label{eq: inflaton value 2}\end{equation}

This result represents a lower bound for the value of the Inflaton
field that arises from the LSC data on the RSBGWs and it has to be
applied to the case of the Pre-Big-Bang Theory \cite{key-4,key-16}. 

It is well known that the requirement for inflation, which is $p=-\rho$
\foreignlanguage{italian}{\cite{key-9,key-10}}, can be approximately
met if one requires $\dot{\varphi}<<V(\varphi)$, where $(\varphi)$
is the potential density of the field. This leads to the famous \textit{slow-roll
approximation} (SRA), which provides a natural condition for inflation
to occur \foreignlanguage{italian}{\cite{key-9,key-10}}. The constraint
on $\dot{\varphi}$ is assured by requiring $\ddot{\varphi}$ to be
negligible. With such a requirement, the slow-roll parameters are
defined (in natural units) by \foreignlanguage{italian}{\cite{key-9,key-10}}

\begin{equation}
\begin{array}{c}
\epsilon(\varphi)\equiv\frac{1}{2}(\frac{V'(\varphi)}{V(\varphi)})^{2}\\
\\\eta(\varphi)\equiv\frac{V''(\varphi)}{V(\varphi)}.\end{array}\label{eq: slow-roll}\end{equation}

Then, the SRA requirements are\foreignlanguage{italian}{ \cite{key-9,key-10}}:

\begin{equation}
\begin{array}{c}
\epsilon\ll1\\
\\|\eta|\ll1,\end{array}\label{eq: slow-roll2}\end{equation}

that are satisfied when it is \foreignlanguage{italian}{\cite{key-9,key-10}}

\begin{equation}
\varphi\gg M_{Planck},\label{eq: planckiano}\end{equation}

where the Planck mass, which is $M_{Planck}\simeq2.177*10^{-5}grams$
in ordinary units and $M_{Planck}=1$ in natural units has been introduced\foreignlanguage{italian}{
\cite{key-9,key-10}}.

Then, one sees immediately that the value of the Inflaton field of
eq. (\ref{eq: inflaton value}), that arises from the WMAP bound on
the RSBGWs, is totally in agreement with the slow roll condition on
Inflation. On the other hand, the value of the Inflaton field of eq.
(\ref{eq: inflaton value 2}), that arises from the LSC bound on the
RSBGWs, is of the order of the Planck mass, thus, it could not be
in agreement with the slow roll condition on Inflation.

The fact that the spectrum of the RSBGWs decreases with increasing
Inflaton field is not surprising. In fact, even if the amplification
of zero-point quantum oscillations increases the spatial dimensions
of perturbations, it is well known that the curvature of the Universe
is {}``redshifted'' by Inflation, i.e. the inflationary scenario
`drives' the universe to a flat geometry \foreignlanguage{italian}{\cite{key-9,key-10}}.

\section{Conclusion remarks}

By using a formula that directly connects the average amplitude of
the RSBGWs with the Inflaton field and the equation for the characteristic
amplitude $h_{C}$ for the RSBGWs, in this proceeding the upper bounds
on the RSBGWs from the WMAP and LSC data have been translated in lower
bounds on the Inflaton field.

The results show that the value of the Inflaton field that arises
from the WMAP bound on the RSBGWs is totally in agreement with the
famous slow roll condition on Inflation \foreignlanguage{italian}{\cite{key-9,key-10}},
while the value of the Inflaton field that arises from the LSC bound
on the RSBGWs could not be in agreement with such a condition. 

Finally, we further emphasize the importance of the formula (\ref{eq: fi}).
If the GWs interferometers will detect the RSBGWs in next years, such
a formula will permit to directly compute the amount of Inflation
in the early Universe.

\subsection*{Acknowledgements}

The Associazione Scientifica Galileo Galilei has to be thanked for
supporting this proceeding.


\begin{thebibliography}{19}
\bibitem{key-1}The LIGO Scientific Collaboration, Class. Quant. Grav.
26, 114013 (2009) 

\bibitem{key-2}R. A. Hulse and J. H. Taylor, Astrophys. J. Lett.
195, 151 (1975)

\bibitem{key-3}C. Corda - \textit{\emph{Interferometric detection
of gravitational waves: the definitive test for General Relativity}}
- Honorable Mention Winner at the 2009 Gravity Research Foundation
Awards for Essays on Gravitation, to appear in December 2009 in a
Special Issue of Int. Journ. Mod. Phys. D, pre-print on arXiv:0905.2502v1
{[}gr-qc{]} 15 May 2009\foreignlanguage{italian}{ }

\bibitem{key-4}The LIGO Scientific Collaboration \& The Virgo Collaboration
- An upper limit on the stochastic gravitational-wave background of
cosmological origin - Nature 460, 990-994 (20 August 2009)\foreignlanguage{italian}{ }

\selectlanguage{italian}%
\bibitem{key-5}L.P. Grishchuk - Zh. Eksp. Teor. Fiz. 67, 825 (1974)\foreignlanguage{english}{ }

\selectlanguage{english}%
\bibitem{key-6}A. A. Starobinsky, JETP Lett. 30, 682 (1979) 

\bibitem{key-7}B. Allen - The stochastic gravity-wave background:
sources and detection - in Proceedings of the Les Houches School on
Astrophysical Sources of Gravitational Waves, eds. Jean-Alain Marck
and Jean-Pierre Lasota (Cambridge University Press, Cambridge, England
1998) 

\bibitem{key-8}G. F. Smoot and P.J. Steinhardt - \textit{Gravity's
rainbow} - First Award Winner at the 1993 Gravity Research Foundation
Awards for Essays on Gravitation - Gen. Rel. Grav. 25, 11, 0001-7701
(1993)

\selectlanguage{italian}%
\bibitem{key-9}S. Watson - An Exposition on Inflationary Cosmology-
http://nedwww.ipac.caltech.edu/level5/Watson/Watson\_contents.html
also in http://xxx.lanl.gov/abs/astro-ph/0005003 (2000)

\selectlanguage{english}%
\bibitem[10]{key-10}\foreignlanguage{italian}{D. H. Lyth and A. R.
Liddle - Primordial Density Perturbation, Cambridge University Press
(2009)}

\bibitem[11]{key-11}\foreignlanguage{italian}{C. L. Bennett et al.
- First Year Wilkinson Microwave Anisotropy Probe (WMAP) Observations:
Preliminary Maps and Basic Results - ApJS \textbf{148} 1 (2003)} 

\bibitem[12]{key-12}\foreignlanguage{italian}{S. Bellucci, S. Capozziello,
M. De Laurentis, and V. Faraoni - Position and frequency shifts induced
by massive modes of the gravitational wave background in alternative
gravity - Phys. Rev. D 79, 104004 (2009)} 

\bibitem[13]{key-13}C. Corda - Information on the Inflaton field
from the spectrum of relic gravitational waves - Gen. Rel. Grav. Gen.
Rel. Grav. 42, 5, 1323-1333 (2010)) 

\bibitem[14]{key-14}S. Capozziello, C. Corda and M. F. De Laurentis
- Stochastic background of gravitational waves {}``tuned'' by f(R)
gravity - Mod. Phys. Lett. A 22, 15, 1097-1104 (2007) 

\selectlanguage{italian}%
\bibitem[15]{key-15}\foreignlanguage{english}{K. S. Thorne - Gravitational
radiation - in 300 Years of Gravitation - eds. S.W. Hawking and W.
Israel, Cambridge University Press, Cambridge, 330 (1987)}

\bibitem[16]{key-16}R. Brustein, M. Gasperini, M. Giovannini, G.
Veneziano - Relic Gravitational Waves from String Cosmology - Phys.
Lett. B361 (1995) 45-51 

\bibitem[17]{key-17}C. W. Misner, K. S. Thorne and J. A. Wheeler
- {}``Gravitation'' - W.H.Feeman and Company - 1973\foreignlanguage{english}{ }

\selectlanguage{english}%
\bibitem[18]{key-18}Landau L and Lifsits E - {}``Teoria dei campi''
- Editori riuniti edition III (1999) 

\bibitem[19]{key-19}Wald RM - \textit{General Relativity -} The Universiy
Chicago Press, Chicago (1984) 
\end{thebibliography}
\end{document}